# The search of the stellar clusters in vicinity of YSOs with high and middle masses

**Naira M. Azatyan, Elena H. Nikoghosyan**

NAS RA V. Ambartsumian Byurakan Astrophysical Observatory (BAO),
Byurakan 0213, Aragatzotn province, Armenia
E-mail: nayazatyan@gmail.com

**Abstract.** The results of the searching on the bases of GPS UKIDSS survey's data of dense compact stellar clusters in the vicinity of 20YSOs with high and middle masses are presented. Totally, we have revealed clusters in 13 areas. Around 5 objects (IRAS 18151–1208, IRAS 18316–0602, IRAS 19110+1045, IRAS 19213+1723, IRAS 20056+3350) they are newly detected. The radii and stellar density have significant gradient: from 0.2 to 2.7 pc and from 3 to 1000 stars/pc$^2$ respectively.

## 1. Introduction

Nowadays there are a large number of observational data, which indicate that the process of star formation in the stellar clusters of the galactic disk is a multistage process, and therefore, in one cluster can be simultaneously observed the objects at different stages of evolution. Perhaps, the embedded dense compact groups or compact clusters of young stellar objects located around the YSO or a pair of YSOs with high or middle masses can be referred to the youngest objects(Lada& Lada 2003).The study of such young formations may be of interest to a number of issues related to the theory of evolution of both the separate stellar objects with different masses and the clusters in general.

After the development of observational astronomy in infrared and radio ranges have been published a number of works on the search and study of such centers of star formation (Bica et al. 2003, Kumar et al. 2006 - 2MASS, Mercer et al. 2005 - Spitzer GLIMPS).However, it should be noted that search of such compact young clusters does not always give a positive result. For example, in (Kumar et al. 2006) the clusters are revealed only in the vicinity of57 out of 217 YSOs with high and middle masses(only 26%). Why they are not detected in other 74% areas? Is it reflecting the real situation or incompleteness of observational data?

The aim of this work is also the search of the compact clusters associated with active YSOs with high and middle masses based on the deep NIR survey (K < 18.$^m$05) GPS UKIDSS.

## 2. The objects selection and investigation methods

### 2.1. Objects selection.

For the search the compact clusters have been chosen the IRAS sources from the list of objects in Varricatt et al.(2010). This list includes 50 YSOs with high and middle masses, in which the various manifestation of activity ($H_2$ and CO outflows, $NH_3$, $H_2O$ and $CH_3OH$ emission and etc.) are observed. All objects are associated with UC HII regions. From this list have been chosen the objects covered by GPS UKIDSS. In total 20 areas were chosen and one of them, namely, IRAS 05137+3919 was already detail examined in Nikoghosyan&Azatyan (2014), where in cluster with radius 1.5′ the ~80 PMS were revealed.

### 2.2. The used data.

In this work, we used the images, astrometric and photometric data obtained from GPS UKIDSS survey (Lucas et al. 2008). The astrometric accuracy and resolution of images are ~ 0.1arcsec/pix. For the purity of the sample with respect to the photometric parameters, we have selected objects with $K<18.^m05$. Furthermore, we excluded those objects for which the probability that they are the result of various kinds of noise or defects of image is more than 90% and objects, which coordinates coincide with the $H_2$ knots of outflows (Varricatt et al. 2010).

### 2.3. The clusters identification.

To identify the cluster we have constructed the spatial distribution around each IRAS source in the 4′ x 4′ area. The density was determined simply by dividing the number of stars by the square of box with b x b dimension and step a. For each cluster, sizes of box and step determined empirically to improve the statistical significance of local peaks stellar density in order to maximize the ability to detect the clusters. The cluster was considered really exist, if the surface density around IRAS sources exceed the average density of background more than $2\sigma$. The isochrones of areas, where revealed the clusters are presented in Fig. 3(the lowest level of isochrones is higher of average field density on $2\sigma$).

      To confirm the existence of the clusters, as well as to refine their size, we have built also a radial distribution of surface density with respect to the geometric center of the cluster, whose position is marked by cross in Fig. 3. As the radius of the cluster, we have taken the distance from the cluster center, from which, according to the Poisson distribution, the fluctuation of the stellar density in the rings, with a probability of more than 1%, has been random character. One example of the radial density distribution in the cluster is presented in Fig. 1. The density distribution clearly shows that cluster radius is 1.5' and the objects in the cluster distribute not uniform forming subgroups, which also reflect isochrones in Fig. 3.

## 2.4. Photometric analyzes.

For the study of nature of the stellar objects of the clusters, we have built JH/HK two-color diagram. One example of the JH/HK two-color diagram is presented in Fig. 2. The J, H, K magnitudes were corrected for absorption, borrowed from http://astro.kent.ac.uk/~df/query_input.html web page. As we can see on the diagram, the significant number of objects are located to the right from the reddening vectors, fall in the area of the Herbig Ae/Be, TTau stars and Class I YSOs, i.e. have a significant infrared excess and are likely PMS stars candidates. We have estimated the probable number of PMS objects in each cluster(N (PMS), see Table 1).

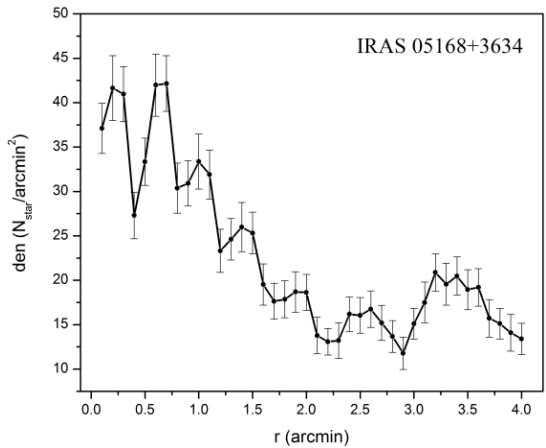

**Fig.1.** The radial distribution of stellar density.

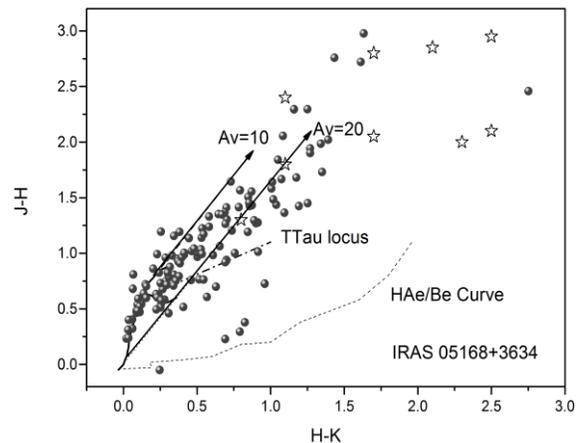

**Fig. 2.** JH/HK two-color diagram. MS and giant branch are adopted from (Bessell& Brett, 1988), TTau locus and reddening vector from (Meyer & Calvet, 1997), H Ae/Be curve and location of Class I YSOs from (Lada & Adams, 1992). To convert JHK magnitudes from UKIRT to CIT photometric system were used the ratios from (Carpenter, 2001).

To conform presence of clusters by KLF we applied Kolmogorov-Smirnov test and compared distributions of magnitudes in the cluster region and surrounding field. Last column in Table 1 represent the results of Kolmogorov-Smirnov test ($P_{KS}$).

## 3. Results.

Using the above methods, we were able to detect the presence of the stellar cluster around 13 out of 19 IRAS sources. The main parameters for 10 of them are presented in Table 1, where are:
(1) - IRAS source;
(2) and (3) - the coordinates of the geometric center
(4) - the distances to the IRAS sources, barrowed from (Varricatt et al. 2010 and ref. therein);
(5) - radiuses of cluster;

(6) - cluster richness (the excessive number of stellar objects relative to field average surface density);
(7) - the number of probable candidates for PMS stars;
(8) - the maximum difference between distributions of magnitudes in cluster and field obtained by KS test.

*Table 1*

## THE PARAMETERS OF THE YOUNG STELLAR CLUSTERS

| IRAS (1) | RA [2000] (2) [h;m;s] | Dec [2000] (3) [°;';"] | d [kpc] (4) | R' (pc) (5) | N (6) | $N_{PMS}$ (7) | $P_{KS}$ (8) |
|---|---|---|---|---|---|---|---|
| 05168+3634 | 05 20 9.2 | +36 37 10 | 6.1 | 1.5 (2.7) | 93 | 80 | 0,003 |
| 05358+3543 | 05 39 11.8 | +35 45 46 | 1.8 | 1.1 (0.6) | 154 | 73 | 0.007 |
| 18151−1208 | 18 17 58.2 | −12 07 30 | 3.0 | 0.2 (0.2) | 7 | 11 | 0.027 |
| 18316−0602 | 18 34 21.3 | −05 59 44 | 3.2 | 0.5 (0.5) | 53 | 38 | 0.001 |
| 18507+0121 | 18 53 17.4 | +01 25 03 | 3.9 | 0.4 (0.5) | 45 | 13 | 0.010 |
| 19110+1045 | 19 13 23.2 | +10 50 58 | 6, 8.3 | 0.3 (0.5, 0.7) | 52 | 53 | 0.469 |
| 19213+1723 | 19 23 37.3 | +17 29 01 | 4.3 | 0.2 (0.25) | 22 | 22 | 0.031 |
| 20056+3350 | 20 07 31.5 | +33 59 39 | 1.7 | 0.6 (0.3) | 93 | 53 | 0.041 |
| 20188+3928 | 20 20 39.3 | +39 38 08 | 3.9 | 0.5 (0.6) | 13 | - | 0.000 |
| 20198+3716 | 20 21 41.1 | +37 25 54 | 0.9, 5.5 | 1.3 (0.3, 2.1) | 385 | 298 | 0.010 |

(1) - IRAS source, (2) и (3) - coordinates of the geometric center, (4) - distance, (5) - cluster radius,
(6) - richness, (7) - number of probable PMS stars, (8) - probability of KS test.

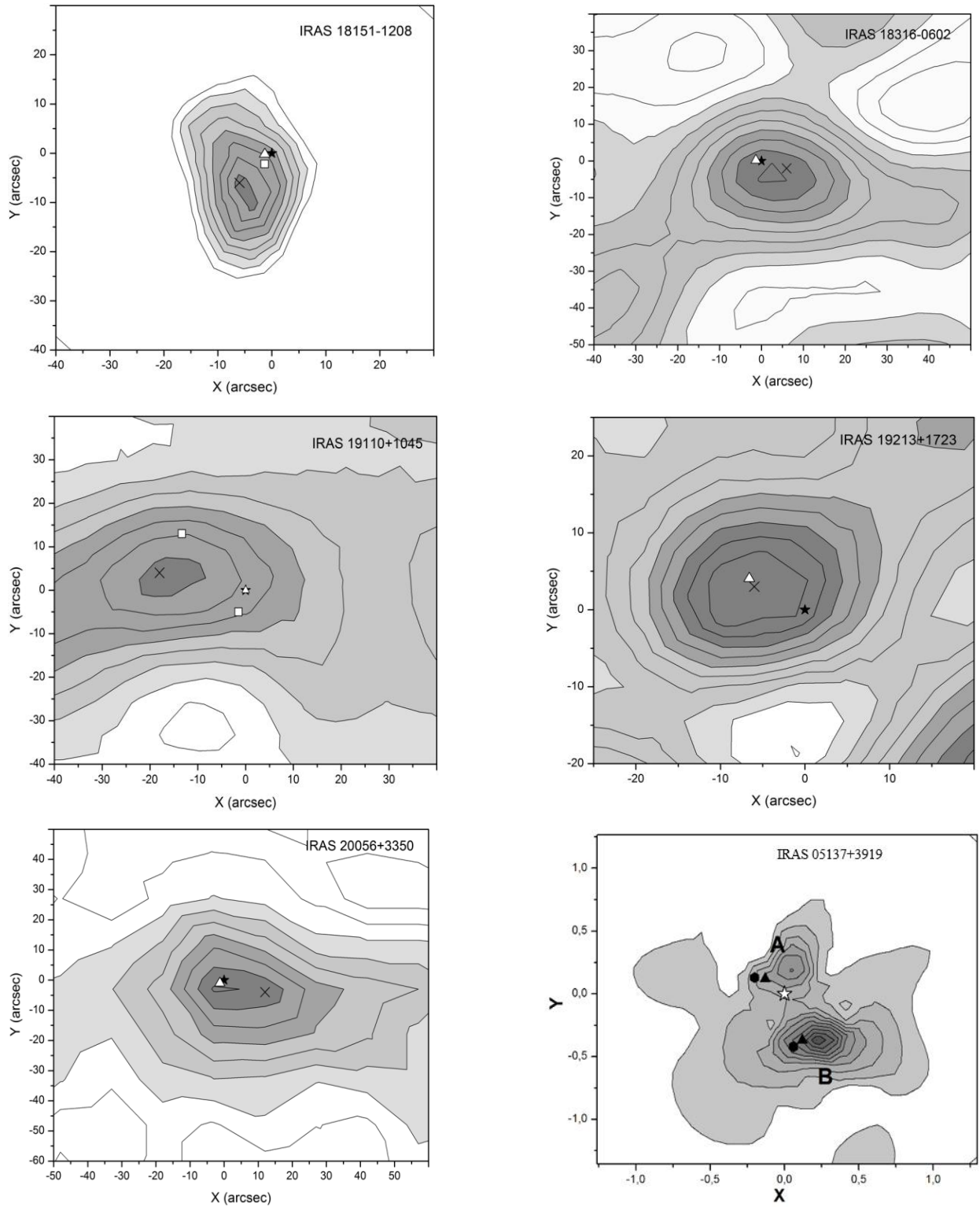

**Fig. 3.** The maps of the surface density of clusters. By asterisks marked the location of IRAS sources, by crosses - the geometric centers, by triangles - MSX sources (Egan et al. 2003), by squares - SCUBA sources (Di Francesco et al. 2008, McCutcheon et al. 1995).

In addition to above mentioned 10 IRAS sources, clusters was found in the vicinity of IRAS19374+2352 and IRAS19388+2357 too. Because for these compact stellar groups(R < 0.5′), by using the same methods, similar results have been already obtained in Faustini et al.(2009) we don't include them in Table 1. In addition, we estimated the number of probable PMS stars ($N_{PMS}$), which is 55 and 32 respectively.

Around of 7 IRAS sources, namely IRAS 18174-1612, IRAS18360-0537, IRAS18385-0512, IRAS18517+0437, IRAS19092+0841, IRAS19410+2336, IRAS20126+4104 were not revealed the stellar clusters in NIR range.

## 4. Conclusion.

Thus, on the bases of GPS UKIDSS survey data from 20 IRAS sources, including IRAS 05137+3919, in the vicinity of 13 of them the compact young stellar clusters were revealed. This amounts 60% of the total study areas, which is much higher than the results obtained according to the 2MASS survey. The radii of clusters and stellar density have significant gradient and are in the range from 0.2 to 2.7 pc and from 3 to 1000 stars/arcmin$^2$ respectively.